\documentclass[review]{elsarticle}
\makeatletter
\def\ps@pprintTitle{%
 \let\@oddhead\@empty
 \let\@evenhead\@empty
 \def\@oddfoot{}%
 \let\@evenfoot\@oddfoot}
\makeatother
\usepackage{lineno,hyperref,graphicx}
\usepackage[normalem]{ulem}
\usepackage[margin=1.0in]{geometry} 
\modulolinenumbers[1]
\setpagewiselinenumbers
\usepackage{multirow}
\usepackage{subfigure}
\usepackage{amsfonts}
\usepackage{amsmath}
\usepackage{booktabs, threeparttable}
\usepackage{array}
\newcolumntype{L}[1]{>{\raggedright\arraybackslash}p{#1}}

\usepackage{color}

\begin{document}

\title{Non-local linear-response functions for thermal transport computed with equilibrium molecular-dynamics simulation}

\author[ucfaddress]{Kevin Fernando}
\author[ucfaddress,AMPACaddress]{Patrick K. Schelling\corref{mycorrespondingauthor}}
\cortext[mycorrespondingauthor]{Corresponding author.}
\ead{patrick.schelling@ucf.edu}
\address[ucfaddress]{Department of Physics, University of Central Florida, Orlando, FL 32816-2385, USA}
\address[AMPACaddress]{Advanced Materials Processing and Analysis Center, University of Central Florida, Orlando, FL 32804, USA}
\begin{abstract}
We establish an approach to compute linear-response functions to elucidate heat waves and non-local thermal transport. The theory is able to describe the response of a system to external heat sources that are nonuniform in space and time. The response functions are computed using equilibrium molecular-dynamics simulations of an Ar crystal modeled using the standard Lennard-Jones potential. It is shown that for low temperatures and short length scales, transport can be partially or even completely ballistic, with the response primarily limited by the group velocity of lattice waves.  By contrast, at longer length scales and higher temperatures, the response functions correspond more closely to diffusive transport characteristic of Fourier's law. It is also shown how the effective thermal conductivity can be determined in a partially-ballistic regime. The results demonstrate the known reduction in the effective thermal conductivity observed when system dimensions are smaller than the mean-free path for lattice waves. Finally, we show how determination of the relevant response functions can be used to model heating of a crystal without requiring additional atomic-scale simulations. Differences between computed results and predictions from Fourier's law represent wave-like, partially-ballistic transport.

\end{abstract}

\maketitle
\section{Introduction}
Thermal conduction is most often described using Fourier's law. Specifically, the Cartesian components of the heat-flux density $\vec{J}$ are related to the temperature gradient $\vec{\nabla}T$ using,
\begin{equation}\label{fourier}
J_{\mu} =- \sum_{\nu} \kappa_{\mu \nu} {\partial T \over \partial x_{\nu}}
\end{equation}
in which $J_{\mu}$ is a component of the heat-flux density, $x_{\nu}$ is a component of the position vector $\vec{x}$, and $\kappa_{\mu \nu}$ is an element of the thermal conductivity tensor. An important assumption of Eq. \ref{fourier} is that the heat-flux density depends only on the local temperature gradient at a single instant of time. For Fourier's law to accurately describe transport, the separation between any heat sources and sinks must be large enough that transport is diffusive. In many instances this is a good assumption, but in nanoscale devices is often not the case \cite{cahill:2003,cahill:2014}. Specifically, if the separation between a heat source and sink is comparable or less than the mean-free path $\Lambda$ of heat carriers, then transport will be at least partially ballistic and transport will deviate significantly from the predictions based on Eq. \ref{fourier}. Moreover, when the heat sources and sinks depend on time, heat transport should be limited by the group velocity of lattice waves. This feature, which is particularly relevant for partially-ballistic transport, is not captured by Fourier's law. It is clear that transport phenomena in a partially-ballistic regime requires a nonlocal theory.

Partially-ballistic transport has been directly observed in ``direct'' molecular-dynamics (MD) simulations which use a heat source and sink and compute the steady-state temperature gradient. It is then assumed that Fourier's law can be applied to the results, with the computed effective thermal conductivity $\kappa_{eff}$ generally differing, often substantially, from the actual bulk thermal conductivity $\kappa$. Specifically, it was shown in Ref. \cite{schelling:2002} that MD simulation yields a linear dependency of ${1 \over \kappa_{eff}}$ on $1 \over L$ where $L$ is the separation between the heat source and sink. Over the past severals years, improved insight and more accurate calculations have demonstrated that the linear dependence assumed in Ref. \cite{schelling:2002} is overly simplistic. This is understood to be due to the fact that scattering rates of vibrational modes depend rather strongly on wavelength \cite{sellan:2010}. Partially-ballistic transport also results in nonlinear temperature profiles \cite{allen:2018}. In Ref. \cite{allen:2018} a nonlocal theory for thermal conduction was developed to probe partially-ballistic effects. Specifically, a non-local function $\kappa(\vec{x},\vec{x}^{\prime})$ was used to relate the temperature profile in an MD calculation to the heat-current density via the expression,
 \begin{equation} \label{nl_allen}
J_{\mu}(\vec{x},t) = - \int \kappa(\vec{x},\vec{x}^{\prime}) {\partial T(\vec{x}^{\prime}) \over \partial x_{\mu}^{\prime}} d^{3}x^{\prime}
\end{equation}
This theory was used in reciprocal space to describe results of extensive MD calculations using the direct method for conduction in GaN \cite{zhou:2009}, with the assumption that for an ideal periodic system $\kappa(\vec{x},\vec{x}^{\prime})=\kappa(\vec{x}-\vec{x}^{\prime})$, the Fourier-transformed response $\tilde{\kappa}(\vec{k})$ was directly obtained. The dependence $\tilde{\kappa}(\vec{k})$ on the magnitude $|\vec{k}|$ was interpreted by solving the Peierls-Boltzmann Equation (PBE) with different models for the dependence of scattering time on phonon wave vector. Solutions were obtained within the relaxation-time approximation (RTA) and very good agreement with MD results was demonstrated \cite{allen:2018}. An important result of this approach was that insight into how scattering rates depend on phonon wave vector was obtained. Moreover, the approach developed in Ref. \cite{allen:2018} demonstrates a way to extract and understand nanoscale transport effects from an MD simulation.

Another simulation approach for obtaining thermal transport properties is to use equilibrium MD simulation and the Green-Kubo (GK) formulation of the fluctuation-dissipation theorem\cite{kubo:1957,green:1954}.  Specifically, the thermal conductivity of a system in equilibrium at temperature $T$ can be computed using,
\begin{equation}\label{gk_standard}
\kappa_{\mu \nu} = {\Omega \over  k_{B}T^{2}} \int_{0}^{\infty} \langle J_{\mu} (\tau) J_{\nu} (0)\rangle d \tau
\end{equation}
in which $\Omega$ is the system volume. Generally, predictions for $\kappa$ obtained with Eq. \ref{gk_standard} do not agree with direct-method calculations due to partially-ballistic transport in the latter case. Hence, it is generally believed that the Green-Kubo approach is more appropriate for determining bulk transport properties. To the best of our knowledge, Green-Kubo methods have not previously been used to extract nonlocal transport properties.

The theory of nonlocal thermal conductivity was first developed in Ref. \cite{mahan:1988}.
In this paper we consider the response function for the heat-flux density $\vec{J}(\vec{x},t)$ which results from an heat pulse which is added at time $t^{\prime}$. The heat pulse results in a local deviation $\Delta u^{(ext)}(\vec{x}^{\prime},t^{\prime})$ from the equilibrium energy density. The heat-flux density after the heat pulse (i.e. for times $t>t^{\prime}$) is given by the expression,
\begin{equation} \label{nl}
J_{\mu}(\vec{x},t) = -{1 \over C_{V}} \int_{\Omega} \sum_{\nu} K_{\mu \nu}(\vec{x}-\vec{x}^{\prime}, t-t^{\prime}) {\partial \Delta u^{(ext)}(\vec{x}^{\prime},t^{\prime})\over  \partial x_{\nu}^{\prime}}  d^{3}x^{\prime} 
\end{equation}
in which $C_V$ is the heat capacity of the volume $\Omega,$ and the gradient of the energy pulse is taken because only a nonuniform pulse will result in heat flow. The primary objective of this paper is to establish how equilibrium MD simulations can be used to determine the nonlocal response function $K_{\mu \nu}(\vec{x}-\vec{x}^{\prime}, t-t^{\prime})$. We also note the assumption of linearity in Eq. \ref{nl}, which is also used in Appendix A to detail the response to a series of external heat pulses.

The approach taken is to recast Eq. \ref{nl} in reciprocal space.
We consider a Fourier expansion of the heat-flux density and the energy deviation. In a periodic system, the nonlocal response function $K_{\mu \nu}(\vec{x}-\vec{x}^{\prime}, t-t^{\prime})$ can also be written as a Fourier series expansion. After integration over the volume $\Omega$, the response can be written for any wave vector $\vec{k}$ as, 
\begin{equation} \label{resp1}
\tilde{J}_{\mu}(\vec{k},t) =-{i \over c_{V}} \sum_{\nu} k_{\nu} \tilde{K}_{\mu \nu}(\vec{k},t-t^{\prime})\tilde{u}^{(ext)}(\vec{k},t^{\prime})
\end{equation}
in which $c_{V}$ is the volumetric heat capacity $c_{V}={C_{V} \over \Omega}$.
In this paper, Eq. \ref{resp1} will be the starting point for deriving the expression used to determine $\tilde{K}_{\mu \nu}(\vec{k},t-t^{\prime})$ from an equilibrium MD simulation. This expression can also be written in terms of the temperature deviation also defined in reciprocal space $\Delta \tilde{T}^{(ext)}(\vec{k},t^{\prime})$,
\begin{equation} \label{resp2}
\tilde{J}_{\mu}(\vec{k},t) =-i \sum_{\nu} k_{\nu} \tilde{K}_{\mu \nu}(\vec{k},t-t^{\prime})\Delta \tilde{T}^{(ext)}(\vec{k},t^{\prime})
\end{equation}
It should be noted that these expressions are in the spirit of the original Kubo paper\cite{kubo:1957} in that the objective is to determine the response of a system to an external perturbation. However, in Ref. \cite{kubo:1957} this is a mechanical perturbation rather than that due to an external heat source. We will discuss this difference in the next section.

It is important to note a difference between Eq. \ref{nl_allen} and Eq. \ref{nl}. In addition to the time-dependence implied in Eq. \ref{nl}, a key distinction is that Eq. \ref{nl} describes the response to an external heat source. By contrast, Eq. \ref{nl_allen} describes a static situation where the external heat sources and sinks are not described, even though they must exist to maintain a static temperature gradient. Therefore, it is important to note that the response function $K_{\mu \nu}(\vec{x}-\vec{x}^{\prime}, t-t^{\prime})$ is not directly comparable to the nonlocal function $\kappa(\vec{x},\vec{x}^{\prime})$ in Eq. \ref{nl_allen}.  In fact, they describe rather different things. In the final section of the paper, we return to address some differences in these two descriptions of nonlocal transport phenomena.

In the next section, the expressions required for the determination of the nonlocal response function are derived. For a reference point, the same response functions are also determined from Fourier's law. These expressions are essentially analogous to the correlation functions used in a Green-Kubo calculation of the thermal conductivity via Eq. \ref{gk_standard}. This is followed by MD simulations results. It is shown how the response functions differ from those obtained using Fourier's law.  The results show that transport is limited by the wave-propagation velocity, and moreover that for low temperatures and short length scales, transport is partially-ballistic.

\section{Linear-response theoretical expressions for non-local heat transport}

Before investigating the non-local response functions, it is helpful to determine the comparable response function from Fourier's law. This is obtained from the well-known Green's function for the heat diffusion equation given here in reciprocal space. For simplicity, we consider an isotropic system with thermal conductivity $\kappa$. We consider the evolution of a temperature deviation $\Delta T(\vec{x},t)$. The thermal-diffusion equation is,
\begin{equation}
{\partial \Delta T(\vec{x},t) \over \partial t} = \alpha \nabla^{2} \Delta T(\vec{x},t)
\end{equation}
in which $\alpha={\kappa \over c_{V}}$ is the thermal diffusivity. Writing this in reciprocal space, and assuming that the initial temperature deviation is due to an external heat source acting at time $t=0$, we obtain the equation for the temperature deviation for times after the pulse $t>0$,
\begin{equation}
\Delta \tilde{T} (\vec{k},t)  = e^{-\alpha k^{2}t}\Delta \tilde{T}^{(ext)} (\vec{k},0) 
\end{equation}
Here $\tilde{T}^{(ext)} (\vec{k},0)$ is due to some external heat source which generates a temperature deviation to a system initially in equilibrium at time $t=0$. Then $\Delta \tilde{T} (\vec{k},t)$ represents the resulting temperature deviation for subsequent times.  Moreover, the $\mu$th component of the heat-flux density is given by,
\begin{equation}
\tilde{J}_{\mu}(\vec{k},t) = -ik_{\mu} \kappa e^{-\alpha k^{2}t}\Delta \tilde{T}^{(ext)}(\vec{k},0)  
\end{equation}
We will only consider systems with cubic symmetry, such that the heat-flux density is directed along $\vec{k}$, and hence we can define,
\begin{equation}
\tilde{J}_{\mu}(\vec{k},t) = \tilde{J}(\vec{k},t){ k_{\mu} \over k}
\end{equation}
in which $k=|\vec{k}|$. Then we can write,
\begin{equation}
\tilde{J}(\vec{k},t) = -ik \kappa e^{-\alpha k^{2}t}\Delta \tilde{T}^{(ext)}(\vec{k},0)  
\end{equation}
In this case the response is also homogenous, and we can identify the response from Fourier's law by comparison to Eq. \ref{resp2},
\begin{equation}
\tilde{K}(\vec{k},t) = \kappa e^{-\alpha k^{2}t}
\label{respdiff}
\end{equation}
This shows that from Fourier's law, an external heat pulse with wave-vector $\vec{k}$ results in a heat-flux density exponentially decays with a characteristic time ${1 \over \alpha k^{2}}$. Also, because the heat-diffusion equation is linear, the response to a series of heat pulses is easily determined using superposition. To model the response to a series of heat pulses, it will be useful to consider the Fourier transform, 
\begin{equation}
\label{KTfourier}
\tilde{K}_{T}(\vec{k},\omega) = \kappa \int_{0}^{\infty}  e^{-\alpha k^{2}t} e^{i \omega t} dt= \kappa  \left [ {\alpha k^{2} + i \omega \over \left (\alpha k^{2}\right)^{2} + \omega^{2}}\right ]
\end{equation}

We now turn to the nonlocal response function. To determine a computational approach using MD simulations, we note that the method in the original Kubo paper\cite{kubo:1957} requires a mechanical perturbation which is added to the Hamiltonian. The added term generates a deviation from equilibrium distribution function. However, it is assumed that the effect is small enough so that the system evolves under the action of the original Hamiltonian without the perturbation. It is then clear that determining how fluctuations in dynamical variables are dissipated within the equilibrium ensemble can be used to determine how the system responds to an external perturbation.

The case of thermal disturbances generates a conceptual difficulty which has been noted by others including by Kubo himself\cite{kubo:1957.2}. Namely, there is no unambiguous way to include a purely thermal disturbance as a perturbation to the  Hamiltonian. Therefore, one encounters immediately a difficulty in attempting to generalize the approach from Ref. \cite{kubo:1957} to derive linear-response functions for thermal transport and many other phenomena. The work by Kubo in Ref. \cite{kubo:1957.2} demonstrates the way forward based on Onsager's generalized reciprocity law\cite{onsager1:1931,onsager2:1931}, which essentially states that any fluctuation in the equilibrium ensemble will decay in time. For a system with some set of macroscopic variables $\alpha_{j}$, it is assumed there is a linear relationship between the thermodynamic driving force and the time derivative of the macroscopic variables. Specifically the evolution is governed by the equation\cite{kubo:1957.2} ,
\begin{equation}
\label{kubo1}
\dot{\alpha}_{j} = \sum_{l} G_{jl} \left(\partial S \over \partial \alpha_{l} \right)
\end{equation}
in which $S$ represents the entropy function and hence $T{\partial S \over \partial \alpha_{l}}$ is a thermodynamic driving force. The objective of Ref. \cite{kubo:1957.2} is to compute the kinetic coefficients $G_{jl}$, and to do this one multiplies Eq. \ref{kubo1} by $\alpha_{m}$ to obtain,
\begin{equation}
\label{kubo2}
G_{jm} = -{1 \over k_{B}} \langle \dot{\alpha}_{j} \alpha_{m} \rangle
\end{equation}
in which the angle brackets represent an equilibrium average. From this point, a Kubo expression for the kinetic coefficients is obtained, and hence the calculation for thermal conductivity $\kappa$ via Eq. \ref{gk_standard} can be demonstrated. In its essential aspects, this approach has been used by Helfand \cite{helfand:1960} for thermal conductivity, with detailed derivations in textbooks \cite{kaviany:2014,evans:1990}. It should be noted that there has been at least one attempt\cite{baroni:2018}  to derive the Green-Kubo expression for thermal conductivity using a mechanical perturbation which ``couples'' to the temperature deviation, and thereby obtain a term which can be added to the Hamiltonian so that the procedure in Ref. \cite{kubo:1957} might be followed. However, it is unclear to us exactly why the mechanical perturbation, which is taken to be the Hamiltonian density, should not itself reflect the temperature deviation. At any rate, Ref. \cite{baroni:2018} demonstrates their approach yields the same Green-Kubo expression for thermal conductivity. Here we do not consider this approach further, but rather use an approach similar to Refs. \cite{kubo:1957.2,helfand:1960,kaviany:2014,evans:1990}. 


The starting point we take is given in Eq. \ref{resp1}. Because we will consider a cubic system, we can assume transport is homogenous. Hence we can take, with the heat pulse input at $t^{\prime}=0$,
\begin{equation}
\tilde{J}(\vec{k},t) = -{i \over c_{V}} k \tilde{K}(\vec{k},t) \tilde{u}^{(ext)}(\vec{k},0)
\label{resp3}
\end{equation}
We next assume that we can determine the response to an external perturbation through a determination of how fluctuations are dissipated in the equilibrium ensemble, thereby following the general prescription by Kubo\cite{kubo:1957.2}. Thus, while the response function will be used to describe the effect of an external heat source, the function itself will be determined using equilibrium MD simulation. We rewrite Eq. \ref{resp3} to describe the dissipation of an instantaneous fluctuation $\tilde{u}(\vec{k},0)$,
\begin{equation}
\tilde{J}(\vec{k},t) = -{i \over c_{V}} k \tilde{K}(\vec{k},t) \tilde{u}(\vec{k},0)
\label{resp5}
\end{equation}
We then multiply both sides of Eq. \ref{resp5} by $ik\tilde{u}(-\vec{k},0)$ and take an equilibrium ensemble average of both sides of the equation, leading to,
 \begin{equation}
 ik \langle  \tilde{J}(\vec{k},t) \tilde{u}(-\vec{k},0)\rangle = {1 \over c_{V} }k^{2} \tilde{K}(\vec{k},t) \langle \tilde{u}(\vec{k},0)\tilde{u}(-\vec{k},0)\rangle
 \end{equation}
 Next we take a derivative with respect to $t$ on both sides and use the continuity equation and time-reversal symmetry. Specifically, we make use of the relation \cite{kubo:1957.2},
 \begin{equation}
\langle  \tilde{J}(\vec{k},t) \tilde{u}(-\vec{k},0)\rangle = \langle  \tilde{J} (\vec{k},0) \tilde{u}(-\vec{k},-t)\rangle
 \end{equation}
 Then taking a derivative with respect to $t$, we finally obtain,
  \begin{equation}
{\partial \tilde{K}(\vec{k},t) \over \partial t} 
= c_{V} {\langle  \tilde{J}(\vec{k},t) \tilde{J}(-\vec{k},0) \rangle   \over \langle \tilde{u}(\vec{k},0)\tilde{u}(-\vec{k},0) \rangle}
\end{equation}
 To obtain an expression for the response function $\tilde{K}(\vec{k},\tau)$, the equation above is integrated,
 \begin{equation}
\tilde{K}(\vec{k},\tau)-\tilde{K}(\vec{k},0)=  
c_{V} {\int_{0}^{\tau}\langle  \tilde{J}(\vec{k},t) \tilde{J}(-\vec{k},0) \rangle  dt \over \langle \tilde{u}(\vec{k},0)\tilde{u}(-\vec{k},0) \rangle}
\label{resp0}
\end{equation}
It is evident that the response function should vanish in the limit $\tau \rightarrow \infty$. In addition, $\tilde{K}(\vec{k},0)$ is also equal to zero. This point is very clear on general terms, but it is also proven in the simulation results since the integral on the right-hand side of Eq. \ref{resp0} is shown to be zero in the limit $\tau \rightarrow \infty$.
Hence the final quantity we obtain is the response function,
 \begin{equation}
 \label{intK}
\tilde{K}(\vec{k},\tau)= 
c_{V} {\int_{0}^{\tau}\langle  \tilde{J}(\vec{k},t) \tilde{J}(-\vec{k},0) \rangle dt \over \langle \tilde{u}(\vec{k},0)\tilde{u}(-\vec{k},0) \rangle}
\end{equation}
This quantity, which can be readily computed by equilibrium MD simulation, governs how a system responds to externally-applied heat pulses. To establish deviations from Fourier's law, calculations based on Eq. \ref{intK} should be compared to the same quantity in Eq. \ref{respdiff} determined using the heat-diffusion equation.
For the denominator in Eq. \ref{intK}, we note that in the long-wavelength limit, the average in the canonical ensemble yields the fluctuation formula for the volumetric heat capacity,
\begin{equation}
  \lim_{\vec{k} \rightarrow 0}\langle \tilde{u}(\vec{k},0)\tilde{u}(-\vec{k},0) \rangle = {c_{V}k_{B}T^{2} \over \Omega}
\end{equation}
However, in the calculations reported here, the constant energy ensemble is used, and the values are determined for finite $\vec{k}$ values. Hence, the term $\langle \tilde{u}(\vec{k},0)\tilde{u}(-\vec{k},0) \rangle$ was computed by time-averaging of the MD simulation results.

This section concludes with the Fourier transform,
\begin{equation}
\tilde{K}_{T}(\vec{k},\omega) = \int_{0}^{\infty} \tilde{K}(\vec{k},\tau) e^{i \omega \tau} d \tau
\label{KTnonlocal}
\end{equation}
The choice of the integration limits accounts for the fact that the response is causal, and hence the response function vanishes for $\tau \le 0$. The expression for $\tilde{K}_{T}(\vec{k},\omega)$ can be directly compared to the result obtained above for Fourier's law in Eq. \ref{KTfourier}. Differences are expected to be due to nonlocal transport phenomena and ``wave-like'' heat transport in MD simulations, in contrast to the purely diffusive behavior predicted by Fourier's law.

Finally, it should be noted that $\tilde{K}(\vec{k},\tau)$ is real, and hence $\tilde{K}(\vec{k},\tau)=\tilde{K}(-\vec{k},\tau)$. It then follows that $\tilde{K}_{T}(\vec{k},\omega)$ is complex and should satisfy the relation $\tilde{K}_{T}^{*}(\vec{k},\omega) = \tilde{K}_{T}(\vec{k},-\omega)$. In the next section, we report the approach and results for computing the response functions. These are compared to those which result from Fourier's law so that the nonlocal nature of the transport can be identified.

\section{Computational approach}
The simulation results were obtained using a Lennard-Jones (LJ) solid in an fcc lattice. Periodic-boundary conditions were applied in all three directions. Although the results could be given in terms of reduced units, we used LJ parameters applicable to the model for LJ argon. The pair-potential was determined from,
\begin{equation}
u(r)= 4 \epsilon \left[\left({\sigma \over r}\right)^{12} -  \left({\sigma \over r}\right)^{6} \right]
\end{equation}
For the LJ parameters, we took $\sigma=3.40 \AA$, and $\epsilon={120K \over k_{B}}$.
A smooth cutoff at a distance $r_{c}=3\sigma$ was used. Specifically, the pair potential $u_{mod}(r)$ given by,
\begin{equation}
u_{mod} (r)  = u(r) - u(r=r_{c}) - \left[ {du \over dr} \right]_{r=r_{c}} \left(r - r_{c}\right)
\end{equation}
was used in the calculations for $r \leq r_{c}$. For $r > r_{c}$,  the pair potential is taken to be zero. It is easily seen that the above expression yields $u_{mod} (r=r_{c})=0$, and moreover the force obtained is also zero at $r=r_{c}$.
Finally, the atomic mass was taken to be $m=39.948$ in atomic-mass units.

The heat-flux density is defined using the standard expression relevant for a pair potential (e.g. see Ref. \cite{schelling:2002} and Ref. \cite{hardy:1963} for a quantum version). Here we distinguish the contribution to the heat-flux density assigned locally to a specific atom. Specifically, we define a component of the heat-flux density associated with atom $i$ as,
\begin{equation}
J_{i,\mu}(t) = {1 \over \Omega} \left[ v_{i,\mu} \left( \epsilon_{i} - \langle h \rangle \right)+ {1 \over 2} \sum_{j \ne i} r_{ij,\mu} \left( \vec{F}_{ij} \cdot \vec{v}_{i}\right)  \right ]
\end{equation}
in which $v_{i,\mu}$ is the $\mu$th component of the velocity $\vec{v}_{i}$, $r_{ij,\mu}$ is the $\mu$th component of the vector $\vec{r}_{ij}=\vec{r}_{i}-\vec{r}_{j}$, and $ \vec{F}_{ij}$ is the pairwise force due to the interaction between atoms $i$ and $j$. The energy $\epsilon_{i}$ is the sum of the potential and kinetic energies, with the potential energy for each pair evenly divided between the two atoms \cite{schelling:2002}. The quantity $\langle h \rangle$ is the average enthalpy per particle, which for a single-component system with $N$ particles is given by $\langle h \rangle = \langle H \rangle/N$, where $H$ is the total system enthalpy. This particular definition with the average enthalpy term subtracted is specifically referred to as the reduced heat-flux density \cite{irving:1950}. The quantity in Eq. \ref{resp2} can be obtained from the above expression by localizing the heat flux at the site of each atom, hence,
\begin{equation}
J_{\mu}(\vec{x},t) = \Omega \sum_{i} J_{i,\mu}(t) \delta^{(3)}(\vec{x}-\vec{r}_{i})
\end{equation}
These definitions are consistent with those derived in Ref. \cite{irving:1950}.
The the Fourier components are given by,
\begin{equation}
\tilde{J}_{\mu}(\vec{k},t)={1 \over \Omega} \int_{\Omega} J_{\mu}(\vec{x},t)  e^{-i \vec{k}\cdot \vec{x}}d^{3}x = \sum_{i} J_{i,\mu} (t) e^{-i \vec{k} \cdot \vec{r}_{i}}
\end{equation}
For $\vec{k}=0$, this reduces to the standard expression of the heat-flux density used in Green-Kubo calculations of the thermal conductivity\cite{schelling:2002}. As mentioned previously, the wave vectors $\vec{k}$ should be chosen to be reciprocal lattice vectors of the simulation supercell, which means that in evaluation of the fourier-transformed heat-flux density does not require any ``minimum image'' convention to be applied.

Two different simulation supercells were used to compute response functions. The small supercell contained $N=4608$ atoms and was orthorhombic with dimensions $3.24$nm $ \times 3.24$ nm $ \times 17.3$nm. The large supercell was also orthorhombic with $N=9216$ atoms and dimensions $3.24$nm $\times 3.24$ nm $\times 34.5$nm. Both supercells correspond to placing the atoms at a nearest neighbor distance $2^{1/6} \sigma$ which corresponds to the minimum of the LJ potential. The larger supercell dimensions were chosen to probe nonlocal transport effects over longer wavelengths while still maintaining manageable computational costs.

The equations of motion were integrated using the velocity Verlet algorithm with MD timestep of $\Delta t= 2.152$fs. Each run was preceded by at least $2.5 \times 10^{4}$ MD steps using a constant temperature algorithm, followed by a much longer run to sample the constant energy ensemble. For $T=12K$ and $T=72K$ using the small supercell, 10 independent simulations were performed with a total simulation time of $5.358ns$ for averaging. For the large supercell, 30 independent calculations at $T=12K$ were performed with a total simulation time $11.384ns$. For simulations at $72K$ in the large supercell, very slow timescales for the response function required longer averaging times. In this case, 40 independent simulations were performed with a total simulation time of $16.677ns$. Error bars for the response functions and Green-Kubo thermal conductivity were determined from analysis of the statistics of the independent runs.

It should be noted that the simulations are entirely classical, and substantially below the Debye temperature of solid Ar of approximately $80$K. However, it has generally been believed that classical simulations can provide insight below the Debye temperature since the most important long-wavelength phonons are described correctly. Generally it is accepted that classical simulations have some validity above the maximum thermal conductivity near $6$K in solid Ar \cite{mcgaughey:2004}. Hence we restrict our calculations to above $10$K

\section{Results}
Before computing the response functions, this section begins with the standard Green-Kubo thermal conductivity $\kappa$.
In Fig. 1, we plot $\kappa(\tau)$ as a function of the upper limit of integration $\tau$ from,
\begin{equation}
 \kappa(\tau)={ \Omega \over k_{B}T^{2}}  \int_{0}^{\tau}  \langle  \tilde{J}(0,\tau) \tilde{J}(0,0) \rangle  d \tau
 \label{kappa}
 \end{equation}
 Formally, the thermal conductivity is equal to the integral in Eq. \ref{kappa} when the integration limit is taken to infinity.  These results provide a solid connection to previous simulation studies of the thermal conductivity of solid Ar. 
 
The results for $\kappa(\tau)$ obtained with Eq. \ref{kappa} in Fig. 1 are plotted for the the supercell with $N=4608$ atoms at $T=12K$.  It can be seen from Fig. 1 that $\tau \approx 40$ps appears to be an adequate integration time to obtain the value of $\kappa$, resulting in $\kappa=1.778 Wm^{-1} K^{-1}$. A longer integration limit $\tau$ result in values $\kappa \approx 2Wm^{-1} K^{-1}$ but with significantly larger error bars. For $T=72K$, the value $\kappa=0.338 Wm^{-1} K^{-1}$ was obtained, again using the integration limit $\tau=40$ps. Computed thermal conductivity values for LJ solids and liquids have been reported elsewhere\cite{tretiakov:2004,mcgaughey:2004,kaburaki:2007,revollo:2018}. We note that our results are comparable in magnitude to these previous works, but exhibit some differences primarily due to differences in the volume per atom and hence the pressure. Specifically, in each of the calculations reported here, the lattice parameter was chosen to be $a=2^{2/3} \sigma$, which corresponds to the nearest-neighbor distance exactly at the minimum of the LJ potential, specifically $2^{1/6}\sigma$. By contrast, previous studies have generally chosen the lattice parameter such that the pressure was zero, $p=0$. These differences result in somewhat lower values for $\kappa$ at low temperatures, but actually higher values of $\kappa$ for high temperatures. It has been verified in earlier studies that $\kappa$ is extremely sensitive to the choice of lattice parameter \cite{mcgaughey:2004,kaburaki:2007}.
 
 The computed values of $\kappa$ can be used within the context of the kinetic theory of heat transport to determine values for the mean free path $\Lambda$. In particular, we assume that $\Lambda = {3\kappa \over C_{V}v_{s}}$.  For lengths $L \sim \Lambda$ and below, it should be expected that the Fourier theory breaks down. From the calculated values of the bulk modulus and the mass density, the sound velocity was determined to be $v_{s}=1110ms^{-1}$.  At the lowest temperature simulated here with $\kappa=1.778Wm^{-1}K^{-1}$, this expression yields $\Lambda \approx 4.5nm$, which is comparable to the cell dimension $L=17.3nm$. This suggests that at least some of the simulation results described  below for finite $\vec{k}$ should correspond to at least partially-ballistic transport. It has also been previously reported that $\Lambda$ tends to depend strongly on mode frequency and wavelength\cite{sellan:2010}. It is likely the case then that at low temperature many modes  have a mean free path $\Lambda$ larger than the simulation supercell.
 
We next turn to the response functions given in Eq. \ref{intK} and Eq. \ref{KTnonlocal}, which represent the main focus of this work.  We first analyze the response functions obtained for the small system with $N=4608$ particles. In Fig. 2, the real part of the response from Eq. \ref{intK} for $T=12K$ is shown as a function of time $\tau$ for a few different wave vectors $\vec{k}$. In particular we consider only nonzero components along the long direction of the supercell from  $k_{z}={2 \pi \over L}$ to the largest value  $k_{z}={8 \pi \over L}$. Some important features emerge. First, as established from the expression in Eq. \ref{intK}, the response function is zero for $\tau=0$. This reflects the fact that the response to any input heat pulse is not instantaneous. In fact, the rise time in Fig. 2 to reach the maximum in the response function is characteristic of the period of a longitudinal acoustic (LA) vibration.  Assuming linear dispersion, which should be approximately true in the long-wavelength limit, this implies for the longest wavelength $\lambda=L=17.3$nm a vibrational period $\tau \approx 15.6$ps. In fact, the maximum response for $k={2 \pi \over L}$ in Fig. 2 occurs at about $\tau = 6.9$ps which is close to half of this vibrational period. Moreover, the response is quite small beyond $\sim 16$ps, indicating that any perturbation is dissipated in a time comparable to the vibrational period. For shorter wavelengths, the response is faster and the time required for the response function to become zero is also shorter. In fact, the times required to reach a maximum are all consistent with half a vibrational period for the relevant LA vibrational mode. 
These facts demonstrate that for $T=12K$, transport over length scales $L \leq 17.3$nm, transport is almost completely ballistic and hence not described by Fourier's law. 

In contrast to the ballistic behavior seen in Fig. 2 at $T=12K$, at $T=72K$ diffusive behavior begins to emerge. The results in Fig. 3 show $\tilde{K}(k,\tau)$ from simulations at $T=72K$.  In each case, the rise time is comparable to the $T=12K$ results in Fig. 2. However, especially in the case $k_{z}={2 \pi \over L}$, the fluctuation persists for a very long time. For $k_{z}={4 \pi \over L}$ the differences between Fig. 2 and Fig. 3 are noticeable but less dramatic. However, for the cases $k_{z}={6 \pi \over L}$ and  $k_{z}={8 \pi \over L}$ which correspond to very short length scales, the differences between Fig. 2 and Fig. 3 are very small, indicating that even at the high temperature $T=72K$ transport is essentially ballistic at these scales.

The real part of the response function $Re \left[\tilde{K}_{T}(\vec{k},\omega) \right]$ is shown in Fig. 4 for $T=12$K and $k_{z}={2 \pi \over L}$. In Fig. 5, the imaginary part $Im \left[\tilde{K}_{T}(\vec{k},\omega) \right]$ is shown. In generating the data, only the real part of $\tilde{K}(\vec{k},\tau)$ was used. As previously noted, $\tilde{K}(\vec{k},\tau)$ is a real function, and any computed imaginary components represent numerical error. For comparison, both Fig. 4 and Fig. 5 show the response functions corresponding to Fourier's theory determined by Eq. \ref{KTfourier} with $\kappa=1.778Wm^{-1}K^{-1}$. Differences between the computed nonlocal response functions and the Fourier theory demonstrate the ballistic nature of the transport. Interestingly, the results show negative values away from the central peak at $\omega=0$, which appear to be primarily due to the finite rise time of the response function. The same quantities are shown for $T=72$K with $k_{z}={2 \pi \over L}$ in Fig. 6 and Fig. 7. The comparison between the Fourier theory obtained with $\kappa=0.338Wm^{-1}K^{-1}$ and the nonlocal response functions show relatively small differences, indicating that transport at this higher temperature is more diffusive in nature. The largest differences between the nonlocal response and Fourier's law are evident near $\omega=0$.

The results for the larger simulation cell with length $L=34.5$nm and $N=9216$ particles show that ballistic behavior persists for $T=12$K, but the response is even more clearly diffusive at $T=72$K in the long wavelength case $k_{z}={2 \pi \over L}$. In Fig. 8, the response function $Re \left[\tilde{K}_{T}(\vec{k},\tau) \right]$ is shown for the large simulation cell at $T=12$K and $T=72L$ for  $k_{z}={2 \pi \over L}$.
 For $T=12$K, the data appears to have a strong ballistic component, but a longer tail beyond $\tau \approx 50$ps which appears to be diffusive. Hence, at these scales for the very low temperature $T=12$K, partially ballistic behavior persists up to $L=34.5$nm. By contrast, at $T=72$K, the results in Fig. 8 appear quite diffusive, with nearly $300$ps of time required to dissipate the initial fluctuation. Comparison of Fig. 3 and Fig. 8, both for $T=72K$ but different system sizes, show that the dissipation time approximately scales as $\sim L^{2}$ which corresponds to diffusive behavior.
 
 In Fig. 9-10 the real and imaginary parts of $\tilde{K}_{T}(\vec{k},\omega)$ are shown for $T=72$K with $L=34.5$nm and $N=9216$ particles. Comparison the response function obtained from Fourier's law demonstrates relatively good agreement, indicating the Fourier's law becomes more suitable at longer length scales and higher temperatures. Hence, the behavior in this case, in agreement with the results in Fig. 8, is more closely associated with diffusive rather than ballistic transport.

Given the response function $\tilde{K}_{T}(\vec{k},\omega)$, the temperature profile can be computed as a response to any external heat source. The details of the calculations are given in Appendix A both for any general response function $\tilde{K}_{T}(\vec{k},\omega)$ and specifically for Fourier's law. Here we specifically consider the response to a very simple sinusoidal heat source given by, 
\begin{equation}
H^{(ext)}(\vec{x},t^{\prime})= {1 \over 4} a \left(e^{i \omega t} + e^{-i \omega t} \right)\left(e^{ i \vec{k} \cdot \vec{x}}+e^{- i \vec{k} \cdot \vec{x}}\right) = a \cos \left( \vec{k} \cdot  \vec{x}\right)\cos \left( \omega t \right)
\label{perturb}
\end{equation}
Here we will consider $\vec{k}={2 \pi \over L} \hat{z}$, where $L$ is the system length along the long direction of the MD simulations described above. For this length, as noted previously, partially-ballistic results are seen at low temperatures, whereas the behavior becomes more consistent with Fourier's law at higher temperatures. The temperature profiles are given by Eq. \ref{TfromKT} in Appendix A when $\tilde{K}_{T}(\vec{k},\omega)$ is obtained from the MD simulations for the small system. For the behavior given by Fourier's law, the temperature profile is described by Eq. \ref{Tfromfourier} also in Appendix A.

The results of this comparison are shown in Figs. 11-13 for an external source with frequency ${\omega \over 2\pi}=0.012$THz at the average temperature $T=12$K. It is assumed in these calculations that the source has been active for a very long period of time so that any transient phenomena can be neglected. For the heating rate we chose a value $a \Omega = 18.1$nW which results in small enough temperature deviations to assume that the linear-response theory  applies. Note that because of the form of the external source in Eq. \ref{perturb}, the external source does not change the overall average system temperature. 

In Fig. 11 and Fig. 12, temperature profiles $T(z,t)$ are shown for the nonlocal theory and Fourier's law assuming $\kappa = 1.778 Wm^{-1}K^{-1}$ corresponding to the average temperature $T=12$K. At  time $t=0$, the external source is adding heat near $z=0$ and $z=L$, and removing heat at $z={L \over 2}$ at a maximum rate.  At $t=0$, the differences between the two theories is rather small as shown in Fig. 11.  By contrast, at the intermediate time, $t={\pi \over 2 \omega}$, when the amplitude of the external source is instantaneously zero, there is a dramatic difference between the two theories, with smaller dissipation in the nonlocal theory. The reason for the difference is due both to a different phase angle $\delta$ and also a different magnitude of the response function $\tilde{K}_{T}(\vec{k},\omega)$ predicted by the two theories. At $T=72$K, the nonlocal theory is expected to predict similar response in comparison to Fourier's theory. In Fig. 13, a comparison is made at time $t={\pi \over 2 \omega}$ where the difference should be most significant.  However, the differences are actually quite small as suggested by the similarity in the response function $\tilde{K}_{T}(\vec{k},\omega)$ at $T=72$K shown in Fig. 6 and Fig. 7. 

In Appendix B, the expressions to determine the effective thermal conductivity $\kappa_{eff}(\vec{k})$ are presented.  The response function $\tilde{K}_{T}(\vec{k},\omega)$ is used to determine  $\kappa_{eff}$ in the static limit of the external source $\omega \rightarrow 0$.  This limit corresponds, for example, to the effective thermal conductivity which is derived from applying the Fourier law to the steady-state temperature profile that is measured or computed in response to a static heat source and sink. In an MD simulation, this corresponds to the usual ``direct'' method of determining $\kappa_{eff}$ \cite{schelling:2002}. It is known that the effective thermal conductivity $\kappa_{eff}$ determined in MD simulations using the ``direct'' method is strongly dependent on system size $L$ when $L \sim \Lambda$, where $\Lambda$ is the mean-free path for heat carriers. In the results reported here, this will be seen as a dependence of $\kappa_{eff}$ on wave vector $k$. Note also that given enough data for very large simulation cells, taking the limit $k \rightarrow 0$ should correspond to the bulk thermal conductivity $\kappa$.

To determine $\kappa_{eff}(k)$, we will consider the smallest frequency ${\omega \over 2\pi}=0.012$THz to approximate the limit $\omega \rightarrow 0$. For $T=12$K and $T=72$K, the results are shown in Fig. 14 as a plot of $1 \over \kappa_{eff}$ as a function of inverse length scale ${k \over 2 \pi}$. Also included in Fig. 14 are the results obtained for the large system with $L=34.5$nm at $k_{z}={2 \pi \over L}$. In addition, the Green-Kubo results from Eq.\ref{gk_standard} are included for $k=0$. The results in Fig. 14 are quite similar to the dependence of the effective thermal conductivity on inverse system size seen in Ref. \cite{schelling:2002} and elsewhere. Qualitatively, the results demonstrate that partially-ballistic transport over short length scales results in smaller effective conductivity values in comparison to the bulk conductivity.

\section{Discussion and conclusions}

The results in this article demonstrate how nonlocal transport properties can be determined using the Green-Kubo approach. Response functions were computed to describe the effective of an external heat source acting on a LJ solid, with comparisons made to Fourier's law to elucidate nonlocal heat transport phenomena, including transport in the partially-ballistic regime.  In Fourier's law, the only relevant quantities are the thermal conductivity $\kappa$ and the volumetric specific heat capacity $c_{V}$ which can be combined to obtain the thermal diffusivity $\alpha = { \kappa \over c_{V}}$.

It was shown how quantitative determination of the response functions $K_{T}(\vec{k},\omega)$ can be used to simulate the response of a system to an external heat source without the need to perform additional MD simulations. Details of this approach are given in Appendix A along with corresponding expressions obtained from Fourier's law. It was shown that at low temperatures and relatively short length scales that transport becomes partially ballistic and Fourier's law predicts substantially different results. 

Finally, it was demonstrated how to obtain values for the effective thermal conductivity $\kappa_{eff}(k)$ corresponding to the response of a system to a static external heat source. As with the effective conductivity obtained using the ``direct'' method, partially-ballistic transport for finite length scales results in effective conductivity values lower than those obtained in the limit $k \rightarrow 0$. Therefore, we have demonstrated how Green-Kubo results for the response functions $K_{T}(\vec{k},\omega)$ in the limit $\omega \rightarrow 0$ can be used to interpret or predict the results of ``direct'' MD simulations of thermal transport. 

We suggest that simulation results for $K_{T}(\vec{k},\omega)$ be used to model the response of systems to external heat sources, especially in regimes where partially ballistic transport is relevant. Specifically, when transport is measured over distances smaller than the mean free path $\Lambda$ and heat waves with transport is primarily limited by the sound velocity can be analyzed and modeled using the approach outlined here. It would also be useful and interesting to compare the response functions obtained from the Green-Kubo methods to predictions from other approaches which go beyond Fourier's law, including the Cattaneo-Vernotte equations \cite{cattaneo:1958,vernotte:1961}, enhanced Fourier's law\cite{ramu:2014}, dual-phase lag model\cite{tzou:1995}, and the mean-free path accumulation function\cite{dames:2005}. It would also be interesting to use this approach to elucidate transport in superlattices, where it has been difficult to connect transport properties to the phonon spectrum of the superlattice in the limit where the mean free path is expected to be larger than the superlattice spacing.

Finally, we return to the question of the differences between Eq. \ref{nl_allen} from Ref. \cite{allen:2018} and Eq. \ref{nl} considered in this paper. In Ref. \cite{allen:2018}, it was demonstrated how the function $\kappa(\vec{x},\vec{x}^{\prime})$ in Eq. \ref{nl_allen} can be determined from analysis of the results obtained from steady-state, nonequilibrium MD simulations. Given those results, it should be possible to determine a response function following the simple approach outlined in Eqs. 7-13. This could be compared to the response function obtained from the assumptions made in Eq. \ref{nl}. It would be very interesting to explore whether these two approaches converge. We expect that in the case of steady-state external sources (i.e. sources with $\omega=0$) the two approaches should converge. However, for time-dependent heat sources (i.e. sources with $\omega \ne 0$), there will be an issue in the application of the nonlocal formulation give by Eq. \ref{nl_allen}. Namely,  Eq. \ref{nl_allen} implies that a heat-flux density arises instantaneously at some distance from where the heat pulse is added to the system. This does not account for the finite wave propagation time. By contrast, the response functions in this paper are time-dependent, and Eq. \ref{nl} is able to account for finite wave propagation speed. 

\section{Acknowledgments}
We acknowledge use of the STOKES cluster provided by the ARCC at UCF. We also would like to acknowledge support from the UCF Burnett Honor's College, and also very useful discussions and inspiration from Prof. Philip Allen.

\section{Data Availability}

The data that support the findings of this study are available from the corresponding author upon reasonable request.

\section{Appendix A}
\setcounter{equation}{0}
\renewcommand{\theequation}{A\arabic{equation}}

In this appendix we derive expressions to show how to compute the heat-flux density and time-dependent temperature profiles due to a time-dependent input power $\tilde{H}^{(ext)}(\vec{k},t^{\prime})$. Starting from Eq. \ref{resp3}, we note that we can write it equivalently with an integral on time,
\begin{equation}
\tilde{J}(\vec{k},t) = -{i \over c_{v}} k \int_{-\infty}^{t} \tilde{K}(\vec{k},t-t^{\prime}) \tilde{u}^{(ext)}(\vec{k},0) \delta(t^{\prime}) dt^{\prime}
\label{resp4}
\end{equation}
The quantity $\tilde{u}^{(ext)}(\vec{k},0) \delta(t^{\prime})$ is an input power, with the energy pulse at $t^{\prime}=0$ given by $ \tilde{u}^{(ext)}(\vec{k},0)$. In the linear regime we can consider a series of pulses due to an input power $\tilde{H}^{(ext)}(\vec{k},t^{\prime})$, and then we have
\begin{equation}
\tilde{J}(\vec{k},t) = -{i \over c_{v}} k \int_{-\infty}^{t} \tilde{K}(\vec{k},t-t^{\prime}) \tilde{H}^{(ext)}(\vec{k},t^{\prime})  dt^{\prime}
\label{resp5}
\end{equation}
In this case the pulse at time $t^{\prime}$ corresponds to $\tilde{H}^{(ext)}(\vec{k},t^{\prime})dt^{\prime}$. In Eq. \ref{perturb} the input power is given in real space, with the reciprocal-space input power defined in the usual way. It is important to note here, as mentioned previously, the assumption of a linear response. This assumption will fail if the amplitude of the external input power becomes large. 

From these considerations we can determine the time-dependent temperature profiles shown in Figs. 11-13 due to the time-dependent heating power given by Eq. \ref{perturb}.
Using the response function $\tilde{K}_{T}(\vec{k},\omega)$,  it is very straightforward to show that the heat-flux density which results from the external perturbation in Eq. \ref{perturb} is given by,
\begin{equation}
\vec{J}(\vec{x},t) = \vec{k}\left(a  \over c_{V}\right)
|\tilde{K}_{T}(\vec{k},\omega)|
\cos \left( \omega t - \delta \right)\sin \left( \vec{k}\cdot \vec{x} \right)
\end{equation}
in which $|\tilde{K}_{T}(\vec{k},\omega)|= \left[\tilde{K}_{T}(\vec{k},-\omega)\tilde{K}_{T}(\vec{k},\omega)  \right]^{1 \over 2}$.
The phase angle $\delta$ is given by,
\begin{equation}
\tan \delta = {Im\left[\tilde{K}_{T}(\vec{k},\omega)\right] \over Re\left[\tilde{K}_{T}(\vec{k},\omega) \right] } 
\end{equation}
Including the external source term, the temperature profile is given by,
\begin{equation}
T(\vec{x},t) = T_{0}  - \left({a  \over c_{V} \omega}\right) \left[
\left({ k^{2}  \over c_{V}}\right) |\tilde{K}_{T}(\vec{k},\omega)|\sin \left( \omega t - \delta \right)
- \sin \left(\omega t \right)\right] \cos \left(\vec{k} \cdot \vec{x}\right)
\label{TfromKT}
\end{equation}
in which $T_{0}$ is the average temperature of the system.
Using numerical results for the response function $\tilde{K}_{T}(\vec{k},\omega)$, the temperature profile can easily be computed as a function of time. Note that there is an implicit assumption that the external perturbation has been active for a very long period of time such that any transient behavior can be ignored.

Given the response function in Eq. \ref{KTfourier} determined using Fourier's law, the external perturbation in Eq. \ref{perturb} results in the heat-flux density,
\begin{equation}
\vec{J}(\vec{x},t)=\vec{k} \left({a  \kappa \over c_{V} }\right){ \cos \left( \omega t - \delta \right)\over \left[(\alpha k^{2})^{2} + \omega^{2} \right]^{1 \over 2}} \sin \left( \vec{k}\cdot \vec{x} \right)
\end{equation}
with the relative phase angle given by,
\begin{equation}
\tan \delta = { \omega \over \alpha k^{2}}
\end{equation}
In this case, the temperature profile is given by,
\begin{equation}
T(\vec{x},t) = T_{0}  - \left({a  \over c_{V} \omega}\right) \left[ \left({\kappa  \over c_{V}}\right) { k^{2}\sin \left( \omega t - \delta \right)\over \left[ (\alpha k^{2})^{2} + \omega^{2}\right]^{1 \over 2}} - \sin\left(\omega t\right)\right]  \cos \left(\vec{k} \cdot \vec{x}\right)
\label{Tfromfourier}
\end{equation}

\section{Appendix B}
\setcounter{equation}{0}
\renewcommand{\theequation}{B\arabic{equation}}

Here we determine the appropriate calculation to obtain the effective thermal conductivity $\kappa_{eff}(\vec{k})$ from the nonlocal response functions.
From the response function in Eq. \ref{KTfourier} obtained from Fourier's law, it can be verified from the expressions for $\vec{J}(\vec{x},t)$ and $T(\vec{x},t)$ that the thermal conductivity $\kappa$ can be determined from the response functions for any wave vector $\vec{k}$ and frequency $\omega$ using,
\begin{equation}
\kappa= \left( {c_{V} \over  k^{2}} \right) {Re\left[ \omega \tilde{K}_{T}(\vec{k},\omega)\right]  \over Im\left[\tilde{K}_{T}(\vec{k},\omega)\right]}
\label{keff}
\end{equation}
For Fourier's law, this expression is exactly correct for any wave-vector $\vec{k}$ and frequency $\omega$. In the more general case with nonlocal response functions, the effective thermal conductivity $\kappa_{eff}(\vec{k})$ is defined in the same way, but requires taking the limit to $\omega \rightarrow 0$,
\begin{equation}
\kappa_{eff}(\vec{k}) = \left({c_{V} \over k^{2}} \right)\lim_{\omega \rightarrow 0} {Re\left[ \omega \tilde{K}_{T}(\vec{k},\omega)\right]  \over Im\left[\tilde{K}_{T}(\vec{k},\omega)\right]}
\label{keff}
\end{equation}
in which the response function corresponds to the nonlocal case determined from MD simulation.
The thermal conductivity $\kappa$ can be approximately determined then from taking the limit of this expression to $\vec{k}=0$, namely $\kappa = \lim_{\vec{k} \rightarrow 0} \kappa_{eff}(\vec{k})$.
For the results shown in Fig. 14, we have used $c_{V}={C_{V} \over \Omega}$ and $C_{V} = 3Nk_{B}$  is the constant-volume heat capacity.

\newpage
      
\begin{center}
{\bf Figure Captions}
\end{center}

\noindent{{\bf Figure 1} Green-Kubo integral from Eq. \ref{kappa} as a function of the upper limit of integration $\tau$ for $T=12K$. The error bars were determined from statistical analysis of 10 independent simulations as described in the text. This result is for the system with $N=4608$ atoms.
}

\noindent{{\bf Figure 2} Real part of the integral in Eq. \ref{intK}  for $\tilde{K}(\vec{k},\tau)$ at temperature $T=12K$ as a function of integration limit $\tau$ for various values of $\vec{k}=k_{z}\hat{z}$. Error bars not shown in instances where they are smaller than the symbols.
}

\noindent{{\bf Figure 3} Real part of the integral in Eq. \ref{intK}   for $\tilde{K}(\vec{k},\tau)$ at temperature $T=72K$ as a function of integration limit $\tau$ for various values of $\vec{k}=k_{z}\hat{z}$. Error bars not shown in instances where they are smaller than the symbols.
}

\noindent{{\bf Figure 4} Real part of the Fourier transform $\tilde{K}_{T}(\vec{k},\omega)$ from Eq. \ref{KTnonlocal} at $T=12K$  as a function of frequency ${\omega \over 2\pi}$ for $k_{z}={2 \pi \over L}$. Comparison is made to the response function given by Fourier's law.
}

\noindent{{\bf Figure 5} Imaginary part of the Fourier transform $\tilde{K}_{T}(\vec{k},\omega)$ from Eq. \ref{KTnonlocal} at $T=12K$  as a function of frequency ${\omega \over 2\pi}$ for $k_{z}={2 \pi \over L}$. Comparison is made to the response function given by Fourier's law.
}

\noindent{{\bf Figure 6} Real part of the Fourier transform $\tilde{K}_{T}(\vec{k},\omega)$ from Eq. \ref{KTnonlocal}  at $T=72K$ as a function of frequency ${\omega \over 2\pi}$ for $k_{z}={2 \pi \over L}$. Comparison is made to the response function given by Fourier's law.
}

\noindent{{\bf Figure 7} Imaginary part of the Fourier transform $\tilde{K}_{T}(\vec{k},\omega)$ from Eq. \ref{KTnonlocal} at $T=72K$  as a function of frequency  ${\omega \over 2\pi}$ for $k_{z}={2 \pi \over L}$. Comparison is made to the response function given by Fourier's law.
}

\noindent{{\bf Figure 8} Real part of the integral in Eq. \ref{intK}  for $\tilde{K}(\vec{k},\tau)$ at temperature $T=12K$ and $T=72K$ as a function of $\tau$ for $k_{z}={2 \pi \over L}$. These results were obtained for the large system size with $L=34.5$nm and $N=9216$ particles.
}

\noindent{{\bf Figure 9} Real part of $\tilde{K}_{T}(\vec{k},\omega)$ from the nonlocal theory compared with the same quantity from Fourier's law for $T=72K$ with $k_{z}={2 \pi \over L}$. These results were obtained for the larger system size with $L=34.5$nm and $N=9216$ particles.
}

\noindent{{\bf Figure 10} Imaginary part of $\tilde{K}_{T}(\vec{k},\omega)$ from the nonlocal theory compared with the same quantity from Fourier's law for $T=72K$ with $k_{z}={2 \pi \over L}$. These results were obtained for the larger system size with $L=34.5$nm and $N=9216$ particles.
}

\noindent{{\bf Figure 11} 
Temperature profiles obtained from the nonlocal response function (solid red line) and Fourier's law (dashed green line) for an input heat source given by Eq. \ref{perturb}. Results shown for time $t=0$ with average temperature $T=12$K.
}

\noindent{{\bf Figure 12} 
Temperature profiles obtained from the nonlocal response function (solid red line) and Fourier's law (dashed green line) for an input heat source given by Eq. \ref{perturb}. Results shown for time $t={\pi \over 2 \omega}$ with average temperature $T=12$K.
}

\noindent{{\bf Figure 13} 
Temperature profiles obtained from the nonlocal response function (solid red line) and Fourier's law (dashed green line) for an input heat source given by Eq. \ref{perturb}. Results shown for time $t={\pi \over 2 \omega}$ with average temperature $T=72$K.
}

\noindent{{\bf Figure 14} Inverse effective conductivity $\kappa_{eff}^{-1}$ as a function of inverse length ${k \over 2 \pi}$ in units $\sigma^{-1}$ obtained from the nonlocal simulation results and Eq. \ref{keff}. Values for $k=0$ obtained from the standard Green-Kubo equation for $\kappa$.
}

\newpage

\end{document}